\documentclass{nime-alternate} 

\usepackage{anonymize}  

\usepackage[utf8]{inputenc}

\begin{document}

\conferenceinfo{NIME'20,}{July 21-25, 2020, Royal Birmingham Conservatoire, ~~~~~~~~~~~~ Birmingham City University, Birmingham, United Kingdom.}
\title{EXPANDING ACCESS TO MUSIC TECHNOLOGY \protect\\Rapid Prototyping Accessible Instrument Solutions For Musicians With Intellectual Disabilities}

%
%
%
\label{key}
%
\numberofauthors{4} 
%
\author{
%
%
\alignauthor
\anonymize{Quinn Jarvis-Holland}\\
       \affaddr{\anonymize{PCC Adaptive \protect\\Instruments Project}}\\
       \affaddr{\anonymize{705 N Killingsworth St}}\\
       \affaddr{\anonymize{Portland, OR 97217}}\\
       \email{\anonymize{quinnjarvisholland@gmail.com}}
\alignauthor
\anonymize{Crystal Cortez}\\
       \affaddr{\anonymize{PCC Adaptive \protect\\Instruments Project}}\\
       \affaddr{\anonymize{705 N Killingsworth St}}\\
       \affaddr{\anonymize{Portland, OR 97217}}\\
       \email{\anonymize{crystalquartez@gmail.com}}
\alignauthor 
\anonymize{Station\titlenote{\anonymize{N. Gammill}}}\\
       \affaddr{\anonymize{PCC Adaptive \protect\\Instruments Project}}\\
       \affaddr{\anonymize{705 N Killingsworth St}}\\
       \affaddr{\anonymize{Portland, OR 97217}}\\
       \email{l\anonymize{nathan.gammill@pcc.edu}}
\and  
\alignauthor 
\anonymize{Francisco Botello}\\
       \affaddr{\anonymize{PCC Adaptive \protect\\Instruments Project}}\\
       \affaddr{\anonymize{705 N Killingsworth St, }}\\
       \affaddr{\anonymize{Portland, OR 97217}}\\
       \email{\anonymize{franciscobotelloungson@gmail.com}}
}
\maketitle
\begin{abstract}
Using open-source and creative coding frameworks, a team of artist-engineers from Portland Community College working with artists who experience Intellectual/Developmental disabilities prototyped an ensemble of adapted instruments and synthesizers that facilitate real-time in-key collaboration. The instruments employ a variety of sensors, sending the resulting musical controls to software sound generators via MIDI. Careful consideration was given to the balance between freedom of expression, and curating the possible sonic outcomes as adaptation. Evaluation of adapted instrument design may differ greatly from frameworks for evaluating traditional instruments or products intended for mass-market, though the results of such focused and individualised design have a variety of possible applications.
\end{abstract}


\section{Introduction}
Innovative Digital Musical Instruments (DMIs) and audio effects can often bypass some presumed necessary abilities or talents. For example, autotuning algorithms allow untrained vocalists to hit perfect pitches and “all in one” mixing effects hide a variety of complex DSP tasks behind a single “macro” knob. 
New technology that is widely accessible consistently gains traction and constantly reshapes the professional landscape of music. Increasingly, even highly technical professions such as mastering have been challenged by AI assisted  DSP.\cite{1}

To remain hopeful in this fast changing climate it is important that we harness and celebrate the ways that these technologies increase access to, and facilitate hyper-localized design and musical expression for under-served populations including people with intellectual/ developmental disabilities.

The idea for a project building adapted instruments was brought forth by Daniel Rolnik (an outsider art critic and curator, who at the time was working with Portland Art and Learning Studios \cite{2}) and Dan Wenger\footnote{PCC Cascade Campus Arts \& Professions Dean} at PCC.  The Portland Community College Cascade campus interactivity lab\footnote{part of the Music and Sonic Arts program \cite{3}}  provides an audio-oriented makerspace and classes teaching microcontrollers and programming to musicians. In contrast to more traditional engineering environments, the lab has empowered many queer, trans, and neuro-divergent artists to utilize electronics and programming. The group of engineers - experienced students and lab techs -all employ a variety of code, sensors, microcontrollers, and midi devices in their own respective installation art and musical performances. 

In the initial meeting Mx Rolnik discussed possible outcomes for the project as well as the needs and current equipment of the artists with cognitive disabilities at PALS. Many artists at the Studios enjoyed making music, but due to the layout of the Studios and the traditional non-adapted instruments (piano, guitar, drumkit) at their disposal, “jamming” could become cacophonous. It was suggested that the instruments in question could somehow connect to each other as artists frequently collaborated in the open-floor Studios space. For more consonant improvisation the designs could employ adaptive constraints to pitch and timing. Mx Rolnik also gave the team a sense of the ways some artists at PALS prefer expressing themselves (eg, vocally-loud, vocally quiet, small sign language, large gestures). Notably, different individuals with the same diagnosis had vastly different preferred range of movements and methods of self expression. This observation is at odds with design processes that rely on broader user-demographic data or designing for all people with a specific intellectual disability.
The initial meeting gave the team a chance to learn important information relating to disability rights, people first language, and the social model of disability-

A social model of disability presents the individual as needing access to support and assistance in the same way as any other person. For example, in a lecture hall chairs are provided for the audience to sit. These are not considered aids or specialist devices; they are normalised as part of the expectation that people would become tired if they had to stand up for any length of time. Chairs are provided for the comfort and assistance of the audience in a completely unremarked way. Yet if someone needs special access to a room or venue this can sometimes be presented as a difficulty and problematised. The principles of inclusive design recommend that it is not only people who walk into a venue who should be considered but all members of society including wheelchair users and parents with prams/buggies.~ \cite{4}
The parallels that can be drawn to considerations for DMI’s are rather clear. In order to best care for people in our communities who experience disabilities we must think of them in the design of things that will be in use in their community. Failing to make accessible designs, according to the social model, is equivalent to building in impairment.
“A person might be very musical and enjoy the violin, saxophone, or piano. But if the individual does not have the use of both arms, he or she is unable to play these instruments. And like buildings, instruments are created by human artisans. So, from a social model perspective, because these instruments are not designed with a person with an impairment in mind, they prevent some individuals from making music with them even though their impairments do not inhibit them from enjoying music.” ~ \cite{5}
Abramo makes an excellent point here applying the social model of disability to instrument design. However the example given doesn’t encompass the variety of experiences and impairments faced by people with disabilities, particularly those experiencing intellectual disabilities. Abramo’s example does not challenge the prevalence of “meritocracy” and a marked penchant for valuing skill or virtuosity above self-expression. Intellectual disabilities do not inhibit people who experience them from enjoying music, or desiring to play music. Aspects that may be inhibiting are having a vast number of controls, requiring precision rhythmic timing, reading quickly, or having too many pitches to choose from. Coincidentally, these are also aspects of an instrument’s design that inhibit non-musicians, children, and the elderly. In turn- building interfaces with the intent of avoiding impairments can serve multiple populations.

\section{Project Parameters}
The team was given a hardware parts budget of 400 USD to build the control surfaces and to design a standalone application(s) in Cycling 74’s Max 8. Keeping a relatively small budget for developing four instruments, and designing freely redistributable applications were both important considerations in order for the project to possibly seed similar efforts at other makerspaces or studios. The most important project parameters were decided on as follows:
\begin{itemize}
    \item 400 budget for parts and materials
    \item Four or more controllers connected to their respective sound generating softwares
    \item Utilize a variety of sensors
    \item Empower artists artists with musical expression, choice
    \item Intuitive and sonically inspiring for skilled or non-disabled musicians
    \item Consonant or in-key improvisation
    \item Facilitate some form of quantized rhythm as well as real-time playing
    \item Sounds complement each other when played together
    \item Low-cost materials
    \item Documentation of designs for DIY re-use
\end{itemize}

Control surfaces were developed using PJRC’s Teensy and Adafruit’s Feather uControllers with the arduino IDE. Both microcontrollers have excellent diagrams, open source schematics, libraries, and other resources online. In particular PJRC’s website has extensive documentation and guides on using usbMIDI in projects. For instruments that benefit from being wireless, the Feather was the preferred dev board due to the built-in (and simple to program) atwinc WIFI chip.
MAX/msp was chosen for developing the sound generation side of the instruments- in part due to the engineers’ familiarity but also because of the immediacy of prototyping with visual and object oriented programming. Routing patches in MAX saves time compared to writing the same DSP operations in openFrameworks, faust, or other lower level languages. Having visual representations of operations and signal flow is particularly useful for members of the team who are visual artists and visual learners. While not open source, MAX can publish executables of a patch, allowing for free distribution of our completed programs. While this has worked for the project parameters with PALS, fully open source alternatives should be explored for further development of the software.

\section{Prototypes And Client Interaction}
The team visited the studio space at PALS to meet some of the artists as well as staff members who were excited about getting new instruments. This served as a time to brainstorm possible form factors and to observe the challenges they currently face with non-adapted instruments. The team spoke with artists there as well as staff facilitators who all had a variety of perspectives and ideas about new instrument solutions.
Using the parameters given at the beginning of the project, communicating with artists and staff clients, and making our own observations, prototype controllers using various sensors were developed  for the clients to play and for the engineers to get a better sense of what kinds of movement and interaction the clients prefer to use. These controllers, paired with sound generation patches in software, quickly found their place as DMI’s.

\subsection{Initial Designs}
\textit{Auto Scaling Touch Synth}\protect\\
A box with small stripes of etched copper on a PCB panel that sends note-on messages when touched. The sound generation is done in a MAX patch with a simple Karplus/Strong implementation to give a plucked string sound.\protect\\
\textit{Interactive Drum Sequencer}\protect\\
A GUI and set of sensor pads that control a drum machine, tempo, and modulation. The software GUI shows accompanying visual feedback of the 4 sensor pads and shows when they have activated one or committed to a sequence. The drum sounds were samples triggered in MAX that would be affected more intensely or less intensely with filters depending on how hard the pad was pressed. With a computer vision script, the computer camera tracked bodies moving to the left or right to speed up or slow down the sequence. This was represented by a slider on the screen as well as shown in a video of themselves in the camera embedded in the GUI.\protect\\
\textit{Rotation Detecting Headband Synth}\protect\\
An accelerometer embedded in a wearable headband, programmed to track the rotation and movement of the head. Sound generation is done in a MAX with a subtractive synthesizer, letting the controller move the notes of a synth up and down in accordance to the degree of rotation of the head left or right. Moving the head side to side triggered an added harmony, and up and down added more or less reverb to the synth. \protect\\
\textit{Handheld 9 Degrees Of Freedom Controller}\protect\\
Compact and wireless handheld device containing a 9 Degrees of Freedom sensor detecting orientation and rotation of the device. The handheld controller drives the same sound generating patch as the headband synth, but with up and down tilting controlling the pitch in glissando, and left and right tilt controlling filter cutoff.\protect\\
\textit{Xbox Kinect Air Harp}\protect\\
An Xbox Kinect IR camera feed is interpreted in a Processing sketch to detect human bodies. The Processing sketch provides visual feedback from what its camera is capturing, through a colored filter, to the GUI. The rectangular area captured by the IR camera is divided into columns and rows of cells. A detected body part moving from one cell to another triggers a note-on at a specific pitch, sent as OSC to MAX - triggering a guitar-like Karplus Strong patch\protect\\

\subsection{Evaluation \& Observations}
There is a vast amount of study into HCI and DMI evaluation, however most traditional or commercially oriented frameworks do not serve minority populations such as people with intellectual disabilities. For this application the team opted for a mostly unstructured client-instrument interaction time. 
Thought was given to the project goals of musical expression, immediacy, performance, and collaboration. Making calls on whether an instrument performance is bad or good is challenging when working with a population that subverts traditional concepts of intellectual hierarchy. For most instruments, those traditional values are a central focus above accessibility.\protect\\

“[...] there may be many perspectives from which to view the effectiveness of the instruments we build. For most performers, performance on an instrument becomes a means of evaluating how well it functions in the context of live music making, and their measure of success is the response of the audience to their performance. Audiences evaluate performances on the basis of how engaged they feel by what they have seen and heard. When questioned, they are likely to describe good performances as " exciting” "skillful" "musical." Bad performances are "boring," and those which are marred by technical malfunction are often dismissed out of hand. If performance is considered to be a valid means of evaluating a musical instrument, then it follows that, for the field of DMI design, a much broader definition of the term "evaluation" than that typically used in human-computer interaction (HCI) is required to reflect the fact that there are a number of stakeholders involved in the design and evaluation of DMIs.” ~\cite{6}

With instrument prototypes functioning, the team and clients met again to beta test and collect observations. For the testing, the instruments were connected to a mixer and PA and initialized to the same key and scale. 
The engineers then stepped back to observe what sorts of interactions arise naturally. Later the engineers took a more hands-on position with the artists to collect feedback and to explore any intended uses of an instrument the clients had not harnessed yet.
Artists preferred to play all of our instruments collectively. 
While the sound modules were different there wasn’t enough sonic space between the sounds we chose for each synth to be fully distinguishable from other instruments in the room, especially because all sounds are emitted from the same speakers.

The artists had vastly varying levels of comfort with each instrument, as with general expression. The most outgoing and physically energetic artists tended to appreciate the instruments using sensors that captured a lot of movement, such as the air harp or 9DOF controller. Artists who were more physically or socially reserved gravitated to the more predictable and involved instruments - the drum sequencer and touch synth
Notable observations by instrument:

\textit{Auto Scaling Touch Synth}
\begin{itemize}
    \item Thin copper strips are too small an area for artists to easily “return” to a desired pitch
    \item Sonic range of such a small number of keys is too limited
    \item Note-on triggers from the controller, and the string-pluck sound module are momentary, sustained key presses do not yield sustained sound
    \item Very immediate and repeatable action to sound reaction
    \end{itemize}
\textit{Interactive Drum Sequencer}
\begin{itemize}
    \item The artists were very engaged with the visual feedback program and were interested in seeing their own image transformed by the choices they were making while playing the instrument. 
    \item The artists were utilizing the face tracking speed control by moving their bodies left to right but it became obvious that the group of artists preferred collaborative creation with the instrument, and the camera’s computer vision was not programmed to track multiple faces and send erroneous control messages.
    \item Different ranges of dexterity and styles of playing became apparent. Some artists preferred lightly touching the pads while others preferred to push hard on them. The sensors in the pads did not allow for this range of pressure application. 
    \end{itemize}
\textit{Headband Synth}
\begin{itemize}
    \item The artists took to playing this instrument intuitively, however there wasn’t an obvious connection between action and sound reaction aside from turning the head left and right and the parameters they were controlling. 
    \end{itemize}
\textit{Handheld 9DOF Controller}
\begin{itemize}
    \item Most artists were very comfortable coming up and grabbing the instrument. 
    \item Since it was small enough to be comfortably held they were able to really move it around freely and discover all kinds of sounds.
    \end{itemize}
\textit{Kinect Air Harp}
\begin{itemize}
    \item Popular with more physically energetic or expressive artists.
    \item Artists enjoyed visual feedback from the camera feed.
\end{itemize}

\subsection{Revision \& Further Development}
After client interaction and testing the team shared observations, and made revisions where appropriate. Roughly halfway through the process from initial tests to final working prototype, the team hosted another testing session with their in-development instruments. This allowed another round of client observation, and strengthened the iterative feedback loops between the artists and instruments, and the observations and design.
These are the revisions and developments for each instrument: \protect\\
\textit{Auto Scaling Touch Synth}
\begin{itemize}
    \item Pluck sound generation patch layered with a polyphonic multi- oscillator synthesizer with sustain and diverse set of possible timbres (programmed into presets)
    \item Blank PCB panels hand painted and etched to yield organic-edged keys
    \item Keyboard expanded from 12 to 24 keys by wiring the etched PCB panels to two MPR121 cap-touch sensing chips
    \item Added two momentary buttons to cycle the scale and cycle through presets
    \item Enclosure designed to house the electronics of the keyboard, to be milled out of plywood
    \end{itemize}
\textit{Interactive Drum Sequencer}
\begin{itemize}
    \item Replaced the piezo sensors with pressure sensors that could detect a greater range of touch 
    \item Sample drum sounds were replaced with synthesized drum sounds in Max which provided a wider range of parameters to manipulate
    \item Added a button to switch between effected and dry drums
    \item Removed the face tracking component and replaced it with a touch capacitive strip on the sequencer itself to set the rate of the sequence, avoiding multiple player bug
    \item Laser cut an enclosure that could be either a handheld or table top instrument
    \end{itemize}
\textit{Headband Synth}
\begin{itemize}
    \item Added the capability to change the key and scale of the synth to fit with other instruments that might be playing as well. 
    \item Added preset options to provide a range of sounds
    \item Exaggerated the effects associated with the up and down and side to side movement of the head to make interaction clearer
    \item Developed a GUI in Max for artists to adjust presets, scale, and key
    \end{itemize}
\textit{Handheld 9DOF Controller}
\begin{itemize}
   \item Upgraded the 9-DOF sensor to one that allowed for more reliable reading of heading (yaw).
    \item Developed a smaller form factor to allow for it to be attached to the hand and allow for more people to use the instrument. 
    \item Installed a battery and wifi chip to make the device un-tethered
    \end{itemize}
\textit{Kinect Air Harp}
\begin{itemize}
    \item Refined note-on detection to produce fewer clustered triggers
    \item Updated the index of note values to include the same scales and keys as the other instruments
    \item Re-arranged the scale degree placement in the camera’s rectangular grid to a more musically and harmonically rewarding pattern
\end{itemize}
One of the most apparent observations from visiting the PALS Studios was that playing music can be difficult when many people with various non-adapted instruments are playing together in close quarters. To play “in sync” with other musicians can require lightning fast perception and solid timing. Fortunately, computers are well adept at both tasks: our team built a master sync patch in max that sends OSC over wifi, any other instruments plugged into laptops running their app can connect to the same wifi network in order to receive the OSC. The primary OSC message controls the variable that picks a particular scale on the “scaler” pre-build in max. Rigid mapping of each instrument to the same scale does restrict a lot of possible musical choices - however the simple setup seemed effective almost immediately. For additional musical variation, the OSC reception can be turned off to manually select a scale. This is notable for vastly increasing the variety of possible harmonies and intervals intervals between two different modes.

\section{Conclusions}
The instruments after revisions and further development, began to serve their intended purposes well. More notably, the sound generators all compliment each other nicely, with certain instruments providing higher notes and others filling in the lows. If there were no budget or time restraints it would be valuable to repeat the testing and revision phases with various client groups.
With all the instruments locked to the same key “wrong” notes are mostly omitted. While this diminishes an individual player’s choices and freedom, it also serves to empower the player by lessening the fear of failure, and increasing the musical cohesion of the player’s resulting sounds with the group’s. While it’s important to consider the freedom in failure and limitlessness, restrictions may not be opposite to creative expression. As explored by Thor Manussen in “Designing Constraints:”
“Margaret A. Boden defines constraints as one of the fundamental sources for creativity: "[F]ar from being the antithesis of creativity, constraints on thinking are what make it possible. . . . Constraints map out a territory of structural possibilities which can then be explored, and perhaps transformed to give another one" (Boden 1990, p. 95). For Boden, the continuity of cultural constraints constitutes the possibility to evaluate creative work, or to recognize ideas as creative. All cultures are founded on constraints,- they are the rule-sets that maintain dynamic unity.” ~\cite{7}
Magnussen and Boden both make a compelling case for constraints empowering expression and creativity here. In this project the greatest success was found in letting artists with disabilities play the instruments together and the result being a traditionally “in tune” and sonically pleasing tapestry. When evaluated against the primary goals, the instrument ensemble designed through this project was a success as it empowers players with the unity of collaborative music making regardless of ability or musical knowledge. Further, the hyper-localized rapid development process used can be applied to similar design issues for minority populations that traditional design processes and evaluation frameworks may not serve.

\subsection{Links}
Documentation and project files will be made available on github: \url{https://github.com/pccadaptiveinstrumentsteam/PCC-Adaptive-Instruments-Project}.

\section{Acknowledgments}
Thanks to ACM SIGCHI and NIME for allowing us to modify templates they have developed.
Our thanks to Daniel Rolnik, the artists and assistants we worked with at PALS, Dean Wenger, PCC, Jesse Mejia, Darcy O’neal, Paul Stoffregen, and Cycling 74\cite{8}

\section{Ethical Standards}
Funding for the project - 400 USD, came from Cycling 74, the publisher of MAX/MSP\cite{8}. To avoid conflict of interest we have avoided claiming their product as better than another, or comparing various commercial products, but rather share why it was useful to us. 
We also acknowledge that the commercial nature of the software is a drawback and are working on - and encourage others to develop - fully open source platforms. Engineers were compensated for hours spent meeting and building the instruments as PCC “casual hire” at 15 USD an hour.
Mx. Rolnik led a discussion for the team around best practices for interacting with the clients with disabilities without disrespecting, overlooking, or misinterpreting their reactions. Mx. Jarvis-Holland also provided information on “People First Language”\cite{9} and the social model of disability to inform the team’s vocabulary and to encourage a musical rather than medical approach.

%
\bibliographystyle{abbrv}

	\bibliography{nime-references}

\begin{thebibliography}{1}

\bibitem{5}
J.~Abramo.
\newblock Disability in the classroom: Current trends and impacts on music
  education.
\newblock {\em Music Educators Journal}, 99(1):39--45, 2012.
\newblock JSTOR \url{www.jstor.org/stable/41692695}.

\bibitem{8}
Cycling 74.
\newblock \url{https://cycling74.com/}.

\bibitem{1}
J.~Drake.
\newblock Ai and music.
\newblock {\em Sound On Sound}, 15(5):795--825, November 1993.

\bibitem{7}
T.~Magnusson.
\newblock Designing constraints: Composing and performing with digital musical
  systems.
\newblock {\em Computer Music Journal}, 34(4):62--73, November 2010.
\newblock JSTOR \url{www.jstor.org/stable/40962941}.

\bibitem{4}
J.~F. Noone.
\newblock The applications of mainstream music technology to facilitate access
  to creative musical experiences for people with disabilities., June 2018.
\newblock \url{
  https://193.1.102.136/bitstream/handle/10344/7570/Noone_2018_Applications.pdf}.

\bibitem{6}
S.~O'Modhrain.
\newblock A framework for the evaluation of digital musical instruments.
\newblock {\em Computer Music Journal}, 35(1):28--42, 2011.
\newblock JSTOR \url{www.jstor.org/stable/41241705}.

\bibitem{2}
PALS.
\newblock Portland art and learning studio.
\newblock \url{https://portlandartandlearningstudio.com/}.

\bibitem{3}
PCC.
\newblock Creative coding program.
\newblock
  \url{https://www.pcc.edu/programs/music-and-sonic-arts/creative-coding.html}.

\bibitem{9}
K.~Snow.
\newblock Disability is natural, 2009.
\newblock \url{https://www.disabilityisnatural.com}.

\end{thebibliography}



\end{document}